\documentstyle[12pt,axodraw]{article}   
  
\newcommand{\bfm}[1]{\mbox{\boldmath$#1$}}  
\newcommand{\ratio}[2]{\mbox{$#1\over#2$}}  
\newcommand{\re}{\mbox{$\rm e$}}  
\newcommand{\ri}{\mbox{$\rm i$}}   
\newcommand{\tr}{\mbox{$\,\rm Tr$}}  
  
\begin{document}
\baselineskip=17pt
\parskip=5pt

\begin{titlepage} \topmargin=-1.0in \footskip=5in

\title{ 
\begin{flushright} \normalsize 
ISU-HET-99-10 \vspace{0.5em} \\ August 1999 \vspace{5em} \\  
\end{flushright}
\large\bf   
Hyperon Nonleptonic Decays in Chiral Perturbation Theory Reexamined}  

\author{
\normalsize\bf A.~Abd~El-Hady$^{(a, b)}$  and  Jusak~Tandean$^{(c)}$ \\ 
\normalsize\it   
$^{(a)}$ Physics Department, Zagazig University, Zagazig, Egypt \\
\normalsize\it   
$^{(b)}$ International Institute of Theoretical and Applied Physics, \\
\normalsize\it   
Iowa State University, Ames, IA 50011 \\
\normalsize\it   
$^{(c)}$ Department of Physics and Astronomy, Iowa State University, 
Ames, IA 50011}

\date{}  
\maketitle   
  
\begin{abstract}  
We recalculate the leading nonanalytic contributions to the amplitudes 
for hyperon nonleptonic decays in chiral perturbation theory.  
Our results partially disagree with those calculated before,  
and include new terms previously omitted in the P-wave amplitudes. 
Although these modifications are numerically significant, they do not 
change the well-known fact that good agreement with experiment cannot 
be simultaneously achieved using one-loop S- and P-wave amplitudes.  
\end{abstract}
  
\end{titlepage}

\section{Introduction}  
  
A number of papers have been devoted to the study of hyperon   
nonleptonic decays  of the form $\,B\rightarrow B'\pi\,$  within   
the framework of chiral perturbation  theory~($\chi$PT).   
These papers have dealt with both the  $\,|\Delta\bfm{I}|=1/2\,$  
components~\cite{bsw,jenkins,JenMan1,GeoCar,springer,BorHol}  and 
the  $\,|\Delta\bfm{I}|=3/2\,$  components~\cite{HeVal,etv}  of 
the decay amplitudes.   
Calculations of the  dominant  $\,|\Delta\bfm{I}|=1/2\,$  
amplitudes have led to mixed results.  
Specifically, theory can give a~good description of either 
the  S-waves or the  P-waves, but not both simultaneously.

In this paper, we revisit the calculation of the leading nonanalytic   
contributions to the  $\,|\Delta\bfm{I}|=1/2\,$  amplitudes.  
Our results disagree for some of the decay diagrams with those of   
Ref.~\cite{jenkins}, which is the most recent published work with   
the same approach.   
Furthermore, our results include new terms in the P-waves that were 
previously omitted.    
We will show that, even though these modifications are numerically  
important, they do not affect the main conclusions of   
Ref.~\cite{jenkins}.

In  Section~\ref{s2}  we review the basic chiral Lagrangian used for 
our calculation in the heavy-baryon formalism.   
Section~\ref{s3}  contains detailed results of our calculation of 
the leading nonanalytic corrections. 
Finally, in  Section~\ref{s4}  we compare our results with  
experiment and present some discussions.

\section{Chiral Lagrangian\label{s2}}     
  
The chiral Lagrangian that describes the interactions of 
the lowest-lying mesons and baryons is written down in terms of 
the lightest meson-octet, baryon-octet, and baryon-decuplet 
fields~\cite{bsw,JenMan1,books}.  
The meson and baryon octets are collected into  $3\times3$  matrices   
$\phi$  and  $B$,  respectively, and the decuplet fields are 
represented by the Rarita-Schwinger tensor  $T_{abc}^\mu$,  
where the notation here follows that of  Ref.~\cite{etv}.  
The octet bosons enter through the exponential  
$\,\Sigma=\exp({\ri}\phi/f),\,$  where  $f$  is the pion-decay 
constant in the chiral-symmetry limit.
In the heavy-baryon formalism~\cite{JenMan1,JenMan2}, the chiral 
Lagrangian is rewritten in terms of velocity-dependent baryon fields,   
$\,B_v^{}(x)=\re^{{\rm i}m_B^{}\not{v}\,v\cdot x} \, B(x)\,$  and  
$\,T_v^\mu(x)=\re^{{\rm i}m_B^{}\not{v}\,v\cdot x} \, T^\mu(x),\,$  
where  $m_B^{}$  is the baryon-octet mass in the chiral-symmetry 
limit.

For the strong interactions, the leading-order chiral Lagrangian is 
given  by~\cite{JenMan1,JenMan2,JenMan3}
\begin{eqnarray}   \label{L1strong}   
{\cal L}^{\rm s}  &=&  
\ratio{1}{4} f^2\,   
\tr \Bigl( \partial^\mu\Sigma^\dagger\, \partial_\mu\Sigma \Bigr)   
\,+\,  
\tr \Bigl( \bar{B}_v^{} \ri v\cdot {\cal D} B_v^{} \Bigr)    
+ 2 D\, \tr \Bigl( \bar{B}_v^{} S_v^\mu   
 \Bigl\{ {\cal A}_\mu^{}, B_v^{} \Bigr\} \Bigr)     
+ 2 F\, \tr \Bigl( \bar{B}_v^{} S_v^\mu   
 \Bigl[ {\cal A}_\mu^{}, B_v^{} \Bigr] \Bigr)   
\nonumber \\ &&   
-\;    
\bar{T}_v^\mu \ri v\cdot {\cal D} T_{v\mu}^{}  
+ \Delta m\, \bar{T}_v^\mu T_{v\mu}^{}  
+ {\cal C} \Bigl( \bar{T}_v^\mu {\cal A}_\mu^{} B_v^{} 
                   + \bar{B}_v^{} {\cal A}_\mu^{} T_v^\mu \Bigr)    
+ 2{\cal H}\, \bar{T}_v^\mu S_v^{}\cdot{\cal A} T_{v\mu}^{}   \;,       
\end{eqnarray}      
where  $\Delta m$  denotes the mass difference between the decuplet 
and octet baryons in the chiral-symmetry limit,  
$S_v^\mu$  is a velocity-dependent spin operator, 
and additional details can be found in  Ref.~\cite{etv}.  
Explicit breaking of chiral symmetry, to leading order in the mass 
of the strange quark and in the limit  $\,m_u^{}=m_d^{}=0,\,$  is 
introduced via  the  Lagrangian~\cite{jenmass}
\begin{eqnarray}   \label{Lmstrong}
{\cal L}_{m_q^{}}^{\rm s}  &=&    
a\, \tr \Bigl( M\Sigma^\dagger+\Sigma M^\dagger \Bigr) 
\,+\,  
b_D^{}\,   
\tr \Bigl( \bar{B}_v^{} 
\Bigl\{ \xi^\dagger M\xi^\dagger + \xi M^\dagger\xi , 
       B_v^{} \Bigr\} \Bigr)   
+ b_F^{}\,    
 \tr \Bigl( \bar{B}_v^{}  
  \Bigl[ \xi^\dagger M\xi^\dagger + \xi M^\dagger\xi , 
        B_v^{} \Bigr] \Bigr)   
\nonumber \\ &&  
+\; 
\sigma\, \tr \Bigl( M\Sigma^\dagger+\Sigma M^\dagger \Bigr) 
 \tr \Bigl( \bar{B}_v^{} B_v^{}  \Bigr)  
\nonumber \\ && 
+\; 
c\, \bar{T}_v^\mu \Bigl( \xi^\dagger M\xi^\dagger 
                        + \xi M^\dagger\xi \Bigr) T_{v\mu}^{}      
- \tilde{\sigma}\, \tr \Bigl( M\Sigma^\dagger 
                               + \Sigma M^\dagger \Bigr)   
 \bar{T}_v^\mu T_{v\mu}^{}   \;,          
\end{eqnarray}   
where $\,M={\rm diag}(0,0,m_s^{}).\,$  
In this limit, the pion is massless and the  $\eta_8^{}$  mass is 
related to the kaon mass by  
$\,m_{\eta_8^{}}^2=\ratio{4}{3}\, m_K^2.\,$    
Moreover, mass splittings within the baryon octet and decuplet 
occur to linear order in  $m_s^{}$.

As is well known, the weak interactions that generate hyperon 
nonleptonic decays are described by a  $\,|\Delta S|=1\,$   
Hamiltonian that transforms as  $(8_{\rm L}^{},1_{\rm R}^{})$  
$\oplus$  $(27_{\rm L}^{},1_{\rm R}^{})$  under  
SU(3$)_{\rm L}^{}\times$SU(3$)_{\rm R}^{}$  rotations.    
Experimentally, the octet piece dominates the 27-plet piece, as 
indicated by the fact that the  $\,|\Delta\bfm{I}|=1/2\,$  
components of the decay amplitudes are larger than  
the  $\,|\Delta\bfm{I}|=3/2\,$  components by about twenty 
times~\cite{etv,overseth}.  
We shall, therefore, assume in what follows  that the decays are 
completely characterized by the  $(8_{\rm L}^{},1_{\rm R}^{})$,  
$\,|\Delta\bfm{I}|=1/2\,$  interactions.  
The leading-order chiral Lagrangian for such interactions  
is~\cite{bsw,jenkins}     
\begin{eqnarray}   \label{weak}  
{\cal L}^{\rm w}  &=&     
h_D^{}\, \tr \Bigl( 
\bar{B}_v^{} \left\{ \xi^\dagger h \xi\,,\,B_v^{} \right\} \Bigr) 
+ h_F^{}\, \tr \Bigl( 
 \bar{B}_v^{} \left[ \xi^\dagger h \xi\,,\,B_v^{} \right]  \Bigr) 
+ h_C^{}\, \bar{T}_v^\mu\, \xi^\dagger h \xi\, T_{v\mu}^{} 
\nonumber \\ &&  
+\; 
\gamma_8^{} f^2\, \tr \Bigl( h\, \partial_\mu^{} \Sigma\,  
                          \partial^\mu \Sigma^\dagger \Bigr)  
\;+\;  {\rm h.c.}   \;,  
\end{eqnarray}      
where  $h$  is a  $3\times3$  matrix with elements  
$\,h_{ij}^{}=\delta_{i2}^{}\delta_{3j}^{},\,$  the parameters 
$h_{D,F,C}^{}$  will be determined below, and  
$\,\gamma_8^{}\approx 8.0\times 10^{-8}\,$  as extracted 
from  kaon  decays.

\section{Amplitudes\label{s3}}
       
Using the heavy-baryon approach, we express the amplitude for 
the decay  $\,B\rightarrow B^\prime\pi\,$  in the form
\begin{eqnarray}        
\ri {\cal M}^{}_{B_{}\rightarrow B_{}'\pi}   \;=\;  
G_{\rm F}^{} m_{\pi^+}^2\, 
\bar{u}_{B_{}'}^{} \left( 
{\cal A}^{(\rm S)}_{B_{}^{}B_{}'\pi}   
+ 2 k\cdot S_v^{}\, {\cal A}^{(\rm P)}_{B_{}^{}B_{}'\pi} 
\right) u_{B_{}^{}}^{}   \;,  
\end{eqnarray}    
where the superscripts refer to the S- and P-wave contributions
and  $k$  is the outgoing four-momen\-tum of the pion.   
The  $\,|\Delta\bfm{I}|=1/2\,$  amplitudes satisfy 
the isospin relations  
\begin{eqnarray}   
\begin{array}{c}   \displaystyle        
{\cal M}^{}_{\Sigma^+\rightarrow n\pi^+}
- \sqrt{2}\, {\cal M}^{}_{\Sigma^+\rightarrow p\pi^0}
- {\cal M}^{}_{\Sigma^-\rightarrow n\pi^-}  \;=\;  0   \;,  
\vspace{1ex} \\   \displaystyle  
\sqrt{2}\, {\cal M}^{}_{\Lambda\rightarrow n\pi^0}
+ {\cal M}^{}_{\Lambda\rightarrow p\pi^-}  \;=\;  0   \;,
\hspace{2em}   
\sqrt{2}\, {\cal M}^{}_{\Xi^0\rightarrow \Lambda\pi^0}
+ {\cal M}^{}_{\Xi^-\rightarrow \Lambda\pi^-}  \;=\;  0   \;, 
\end{array}   
\end{eqnarray}   
and so only four of them are independent. 
Following  Refs.~\cite{bsw,jenkins},  we choose the four to be  
$\,\Sigma^+\rightarrow n\pi^+,\,$  $\Sigma^-\rightarrow n\pi^-,\,$  
$\Lambda\rightarrow p\pi^-\,$  and  
$\,\Xi^-\rightarrow\Lambda\pi^-.\,$

In calculating the decay amplitudes, we will consider 
the leading-order terms and their one-loop corrections.  
For the loop diagrams, we will adopt the approach taken in 
Ref.~\cite{jenkins} by keeping only calculable terms of  
${\cal O}(m_s^{}\ln m_s^{})$  and  ${\cal O}(m_s^2\ln m_s^{})$.  
The latter,  ${\cal O}(m_s^2\ln m_s^{})$,  terms are formally 
smaller than the former, but are included because, as was argued in  
Ref.~\cite{jenkins},  they arise from graphs proportional to  
$\gamma_8^{}$  in  Eq.~\ref{weak}, whose value is enhanced  
with respect to naive expectation, thereby generating contributions 
comparable to the  ${\cal O}(m_s^{}\ln m_s^{})$  terms, which are  
proportional to  $h_{D,F,C}^{}$.  
At the one-loop level, there are also corrections of  
${\cal O}(m_s^{})$,  but these are not computable due to the many  
counterterms that contribute at the same order. 
To take into account the error associated with neglecting all these 
terms,\footnote{To simplify our calculation and to follow 
Ref.~\cite{jenkins},  we have also neglected  ${\cal O}(m_s^{})$  
terms which are calculable from the heavy-baryon expansion of 
the relativistic Lagrangian~\cite{meissner}, as well as calculable  
${\cal O} \bigl( m_s^{3/2} \bigr) $  contributions (from the one-loop 
diagrams) that we expect to be small.}   
we will incorporate some theoretical uncertainties when comparing 
theory with experiment.     
It is possible to perform a complete one-loop calculation including 
all counterterms, but then one loses predictive power as there are 
more free parameters than data.\footnote{Such a calculation was done  
in  Ref.~\cite{BorHol}, without explicitly including the decuplet 
baryons in the effective theory.}

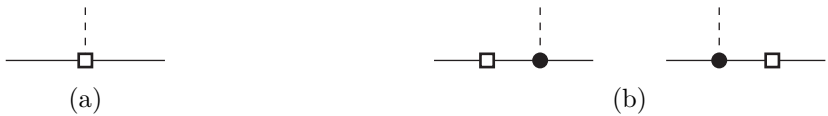
\begin{figure}[ht]         
   \hspace*{\fill} 
\begin{picture}(60,50)(-30,-20)    
\Line(-30,0)(30,0) \DashLine(0,0)(0,20){3} 
\Text(0,-15)[c]{\footnotesize(a)}   
\SetWidth{1}   \BBoxc(0,0)(5,5)         
\end{picture}
   \hspace*{\fill} 
\begin{picture}(60,50)(-30,-20)
\Line(-30,0)(30,0)  
\Vertex(10,0){3} \DashLine(10,0)(10,20){3}    
\SetWidth{1}   \BBoxc(-10,0)(5,5) 
\end{picture}
\begin{picture}(20,50)(-10,-20)
\Text(0,-15)[c]{\footnotesize(b)}   
\end{picture}
\begin{picture}(60,50)(-30,-20)
\Line(-30,0)(30,0)  
\Vertex(-10,0){3} \DashLine(-10,0)(-10,20){3}    
\SetWidth{1}   \BBoxc(10,0)(5,5)          
\end{picture}
   \hspace*{\fill} 
\caption{\label{tree}%
Tree-level diagrams for (a) S-wave and (b) P-wave hyperon 
nonleptonic decays. 
In all figures, a solid (dashed) line denotes a baryon-octet 
(meson-octet) field, and a solid dot (hollow square) represents 
a strong (weak) vertex, with the strong vertices being generated 
by  ${\cal L}^{\rm s}$  in Eq.~(\ref{L1strong}).  
Here the weak vertices come from the  $h_{D,F}^{}$  terms in  
Eq.~(\ref{weak}).}
\end{figure}

We write the S- and P-wave decay amplitudes at the one-loop level in 
the form    
\begin{eqnarray}   \label{As,Ap}  
\begin{array}{c}   \displaystyle  
{\cal A}^{\rm (S)}_{B_{}^{}B_{}'\pi}  \;=\;   
{1\over\sqrt{2}\, f_{\!\pi}^{}} \Biggl[  
\alpha^{\rm (S)}_{B_{}^{}B_{}'}     
\;+\;  
\Bigl( \bar{\beta}^{\rm (S)}_{B_{}^{}B_{}'}     
      - \bar{\lambda}^{}_{B_{}^{}B_{}'\pi} 
       \alpha^{\rm (S)}_{B_{}^{}B_{}'} \Bigl)      
{m_K^2\over 16\pi^2 f_{\!P}^2}\, \ln{m_K^2\over\mu^2}  
\Biggr]   \;,   
\vspace{2ex} \\   \displaystyle   
{\cal A}^{\rm (P)}_{B_{}^{}B_{}'\pi}  \;=\;   
{1\over\sqrt{2}\, f_{\!\pi}^{}} \Biggl[  
\alpha^{\rm (P)}_{B_{}^{}B_{}'}     
\;+\;  
\Bigl( \bar{\beta}^{\rm (P)}_{B_{}^{}B_{}'}     
        - \bar{\lambda}^{}_{B_{}^{}B_{}'\pi} 
         \alpha^{\rm (P)}_{B_{}^{}B_{}'} \Bigl)      
 {m_K^2\over 16\pi^2 f_{\!P}^2}\, \ln{m_K^2\over\mu^2}  
\;+\;  
\tilde{\alpha}^{\rm (P)}_{B_{}^{}B_{}'}   
\Biggr]   \;,
\end{array}    
\end{eqnarray}    
where  $\,f_{\!\pi}^{}\approx 92.4\, \rm MeV\,$  is the physical 
pion-decay constant and  $\,f_{\!P}^{}=f_{\!\pi}^{}$  or  
$f_{K}^{}(\approx 1.22f_{\!\pi}^{})$.    
Contributions from tree-level and one-loop diagrams,
shown in  Figures~\ref{tree},  \ref{s-wave,p-wave,loop},  
and~\ref{g8,loop},  
are represented by  $\alpha^{}_{B_{}^{}B_{}'}$  and  
$\, \bar{\beta}^{}_{B_{}^{}B_{}'} =     
   \beta^{}_{B_{}^{}B_{}'} + \beta'_{B_{}^{}B_{}'},\,$    
respectively,  where  $\beta^{}_{B_{}^{}B_{}'}$  comes from one-loop  
decay graphs involving only octet baryons and  
$\beta'_{B_{}^{}B_{}'}$  arises from those with internal 
decuplet-baryon lines.\footnote{%
In the figures, we have not included a tree-level P-wave diagram 
with a  $\gamma_8^{}$  vertex inserted on the outgoing-meson line 
because it vanishes in the massless-pion limit that we take here.   
We have also not included one-loop P-wave diagrams with three baryon 
propagators inside the loop 
(corresponding to Figures 3l and 3m of  Ref.~\cite{BorHol}), 
which are proportional to  $h_{D,F,C}^{}$  and  of higher  $m_s^{}$  
order,  ${\cal O} \bigl( m_s^{3/2} \bigr) $.}       
The coefficient  $\bar{\lambda}^{}_{B_{}^{}B_{}'\pi}$  contains 
contributions from baryon and pion wave-function renormalization 
as well as the renormalization of the pion-decay constant.  
The P-wave term  $\tilde{\alpha}^{\rm (P)}_{B_{}^{}B_{}'}$  results 
from one-loop corrections to the propagator in the lowest-order 
P-wave diagrams in  Figure~\ref{tree}(b).  
The expressions for  $\alpha_{B_{}^{}B_{}'}^{}$,   
$\beta_{B_{}^{}B_{}'}^{}$,  $\beta'_{B_{}^{}B_{}'}$,  
$\bar{\lambda}^{}_{B_{}^{}B_{}'\pi}$,  and  
$\tilde{\alpha}^{\rm(P)}_{B_{}^{}B_{}'}$  are given in the Appendix.

\begin{figure}[ht]         
   \hspace*{\fill} 
\begin{picture}(80,50)(-40,-20)    
\Line(-30,0)(30,0) \DashLine(0,0)(0,20){3} 
\DashCArc(0,-10)(10,0,360){3}        
\SetWidth{1}   \BBoxc(0,0)(5,5)         
\end{picture}
   \hspace*{\fill} 
\begin{picture}(80,50)(-40,-20)
\Line(-40,0)(40,0)
\DashCArc(0,0)(20,180,360){4} 
\Vertex(20,0){3} \DashLine(20,0)(20,20){3}    
\SetWidth{1}   \BBoxc(-20,0)(5,5) 
\end{picture}
   \hspace*{\fill} 
\begin{picture}(80,50)(-40,-20)
\Line(-40,0)(40,0) 
\DashCArc(0,0)(20,180,360){4} 
\Vertex(-20,0){3} \DashLine(-20,0)(-20,20){3}    
\SetWidth{1}   \BBoxc(20,0)(5,5)          
\end{picture}
   \hspace*{\fill} 
\begin{picture}(80,50)(-40,-20)
\Line(-40,0)(40,0) 
\DashCArc(0,0)(20,180,360){4} 
\Vertex(-20,0){3} \Vertex(20,0){3} 
\DashLine(0,0)(0,20){3}    
\SetWidth{1}   \BBoxc(0,0)(5,5)             
\end{picture}
   \hspace*{\fill} 
\begin{picture}(80,50)(-40,-20)
\Line(-40,0)(-20,0) 
\Line(-20,1.5)(20,1.5) \Line(-20,-1.5)(20,-1.5) 
\Line(20,0)(40,0) 
\DashCArc(0,0)(20,180,360){4} 
\Vertex(-20,0){3} \Vertex(20,0){3} 
\DashLine(0,0)(0,20){3}    
\SetWidth{1}   \BBoxc(0,0)(5,5)             
\end{picture}
   \hspace*{\fill} 
\\   
   \hspace*{\fill} 
\begin{picture}(20,20)(-10,0)
\Text(0,5)[c]{\footnotesize(a)}   
\end{picture}
   \hspace*{\fill} 
\\   
   \hspace*{\fill} 
\begin{picture}(80,50)(-40,-20)    
\Line(-40,0)(-15,0) 
\DashCArc(-15,10)(10,-90,270){3}        
\Line(-15,0)(15,0) 
\Vertex(15,0){3} \DashLine(15,0)(15,20){3} 
\Line(15,0)(40,0) 
\SetWidth{1}   \BBoxc(-15,0)(5,5) 
\end{picture}
   \hspace*{\fill} 
\begin{picture}(80,50)(-40,-20)    
\Line(-40,0)(-15,0) \Vertex(-15,0){3} 
\DashLine(-15,0)(-15,20){3} 
\Line(-15,0)(15,0) 
\DashCArc(15,10)(10,-90,270){3}        
\Line(15,0)(40,0) 
\SetWidth{1}   \BBoxc(15,0)(5,5) 
\end{picture}
   \hspace*{\fill} 
\begin{picture}(80,50)(-40,-20)    
\Line(-40,0)(-15,0) \Vertex(15,0){3} 
\Line(-15,0)(15,0) 
\DashLine(15,0)(15,20){3} 
\DashCArc(15,-10)(10,0,360){3}        
\Line(15,0)(40,0) 
\SetWidth{1}   \BBoxc(-15,0)(5,5) 
\end{picture}
   \hspace*{\fill} 
\begin{picture}(80,50)(-40,-20)    
\Line(-40,0)(-15,0) \Vertex(-15,0){3}     
\DashLine(-15,0)(-15,20){3}   
\DashCArc(-15,-10)(10,-180,180){3}        
\Line(-15,0)(15,0) \Line(15,0)(40,0) 
\SetWidth{1}   \BBoxc(15,0)(5,5)         
\end{picture}               
   \hspace*{\fill} 
\\    
   \hspace*{\fill} 
\begin{picture}(100,50)(-60,-20)
\Line(-60,0)(-40,0) \Line(-40,0)(-20,0)       
\Line(-20,0)(0,0) \Line(0,0)(20,0) \Line(20,0)(40,0) 
\DashCArc(0,0)(20,180,0){4} 
\Vertex(-20,0){3} \Vertex(0,0){3} \Vertex(20,0){3} 
\DashLine(0,0)(0,20){3} 
\SetWidth{1}   \BBoxc(-40,0)(5,5) 
\end{picture}
   \hspace*{\fill} 
\begin{picture}(100,50)(-40,-20)
\Line(-40,0)(-20,0) \Line(-20,0)(0,0) \Line(0,0)(20,0) 
\DashCArc(0,0)(20,180,0){4}        
\Vertex(-20,0){3} \Vertex(0,0){3} \Vertex(20,0){3}         
\DashLine(0,0)(0,20){3} 
\Line(20,0)(40,0) \Line(40,0)(60,0) 
\SetWidth{1}   \BBoxc(40,0)(5,5) 
\end{picture}
   \hspace*{\fill} 
\begin{picture}(100,50)(-60,-20)
\Line(-60,0)(-40,0) \Line(-40,0)(-20,0)       
\Line(-20,1.5)(20,1.5) \Line(-20,-1.5)(20,-1.5) 
\Line(20,0)(40,0) 
\DashCArc(0,0)(20,180,0){4} 
\Vertex(-20,0){3} \Vertex(0,0){3} \Vertex(20,0){3} 
\DashLine(0,0)(0,20){3} 
\SetWidth{1}   \BBoxc(-40,0)(5,5) 
\end{picture}   
   \hspace*{\fill} 
\begin{picture}(100,50)(-40,-20)
\Line(-40,0)(-20,0) 
\Line(-20,1.5)(20,1.5) \Line(-20,-1.5)(20,-1.5) 
\DashCArc(0,0)(20,180,0){4}        
\Vertex(-20,0){3} \Vertex(0,0){3} \Vertex(20,0){3}         
\DashLine(0,0)(0,20){3} 
\Line(20,0)(40,0) \Line(40,0)(60,0) 
\SetWidth{1}   \BBoxc(40,0)(5,5) 
\end{picture}
   \hspace*{\fill} 
\\    
   \hspace*{\fill} 
\begin{picture}(100,50)(-60,-20)
\Line(-60,0)(-40,0) \Line(-40,0)(-20,0)       
\Line(-20,1.5)(0,1.5) \Line(-20,-1.5)(0,-1.5)   
\Line(0,0)(20,0) \Line(20,0)(40,0) 
\DashCArc(0,0)(20,180,0){4} 
\Vertex(-20,0){3} \Vertex(0,0){3} \Vertex(20,0){3} 
\DashLine(0,0)(0,20){3} 
\SetWidth{1}   \BBoxc(-40,0)(5,5) 
\end{picture}
   \hspace*{\fill} 
\begin{picture}(100,50)(-40,-20)
\Line(-40,0)(-20,0) 
\Line(-20,1.5)(0,1.5) \Line(-20,-1.5)(0,-1.5) 
\Line(0,0)(20,0)  \DashCArc(0,0)(20,180,0){4}        
\Vertex(-20,0){3} \Vertex(0,0){3} \Vertex(20,0){3}         
\DashLine(0,0)(0,20){3} 
\Line(20,0)(40,0) \Line(40,0)(60,0) 
\SetWidth{1}   \BBoxc(40,0)(5,5) 
\end{picture}
   \hspace*{\fill} 
\begin{picture}(100,50)(-60,-20)
\Line(-60,0)(-40,0) \Line(-40,0)(-20,0)       
\Line(-20,0)(0,0) \Line(0,1.5)(20,1.5) \Line(0,-1.5)(20,-1.5) 
\Line(20,0)(40,0) 
\DashCArc(0,0)(20,180,0){4} 
\Vertex(-20,0){3} \Vertex(0,0){3} \Vertex(20,0){3} 
\DashLine(0,0)(0,20){3} 
\SetWidth{1}   \BBoxc(-40,0)(5,5) 
\end{picture}
   \hspace*{\fill} 
\begin{picture}(100,50)(-40,-20)
\Line(-40,0)(-20,0) \Line(-20,0)(0,0) 
\Line(0,1.5)(20,1.5) \Line(0,-1.5)(20,-1.5) 
\DashCArc(0,0)(20,180,0){4}        
\Vertex(-20,0){3} \Vertex(0,0){3} \Vertex(20,0){3}         
\DashLine(0,0)(0,20){3} 
\Line(20,0)(40,0) \Line(40,0)(60,0) 
\SetWidth{1}   \BBoxc(40,0)(5,5) 
\end{picture}
   \hspace*{\fill} 
\\  
   \hspace*{\fill} 
\begin{picture}(100,30)(-40,0)
\Line(-40,0)(-20,0) \Line(-20,0)(0,0) \Line(0,0)(20,0) 
\DashCArc(0,0)(20,0,180){4}        
\Vertex(-20,0){3} \Vertex(20,0){3} \Vertex(40,0){3}         
\DashLine(40,0)(40,20){3} 
\Line(20,0)(40,0) \Line(40,0)(60,0) 
\SetWidth{1}   \BBoxc(0,0)(5,5) 
\end{picture}
   \hspace*{\fill} 
\begin{picture}(100,30)(-60,0)
\Line(-60,0)(-40,0) \Line(-40,0)(-20,0)       
\DashLine(-40,0)(-40,20){3} 
\Line(-20,0)(0,0) \Line(0,0)(20,0) \Line(20,0)(40,0) 
\DashCArc(0,0)(20,0,180){4} 
\Vertex(-40,0){3} \Vertex(-20,0){3} \Vertex(20,0){3} 
\SetWidth{1}   \BBoxc(0,0)(5,5) 
\end{picture}
   \hspace*{\fill} 
\begin{picture}(100,30)(-40,0)
\Line(-40,0)(-20,0) 
\Line(-20,1.5)(0,1.5) \Line(-20,-1.5)(0,-1.5) 
\Line(0,1.5)(20,1.5)  \Line(0,-1.5)(20,-1.5)  
\DashCArc(0,0)(20,0,180){4}        
\Vertex(-20,0){3} \Vertex(20,0){3} \Vertex(40,0){3}         
\DashLine(40,0)(40,20){3} 
\Line(20,0)(40,0) \Line(40,0)(60,0) 
\SetWidth{1}   \BBoxc(0,0)(5,5) 
\end{picture}
   \hspace*{\fill} 
\begin{picture}(100,30)(-60,0)
\Line(-60,0)(-40,0) \Line(-40,0)(-20,0)       
\DashLine(-40,0)(-40,20){3} 
\Line(-20,1.5)(0,1.5) \Line(-20,-1.5)(0,-1.5) 
\Line(0,1.5)(20,1.5)  \Line(0,-1.5)(20,-1.5)  
\Line(20,0)(40,0) 
\DashCArc(0,0)(20,0,180){4} 
\Vertex(-40,0){3} \Vertex(-20,0){3} \Vertex(20,0){3} 
\SetWidth{1}   \BBoxc(0,0)(5,5) 
\end{picture}
   \hspace*{\fill} 
\\   
   \hspace*{\fill} 
\begin{picture}(20,20)(-10,0)
\Text(0,5)[c]{\footnotesize(b)}   
\end{picture}   
   \hspace*{\fill} 
\caption{\label{s-wave,p-wave,loop}%
One-loop diagrams contributing to (a) S-wave and (b) P-wave hyperon  
nonleptonic decay amplitudes, with weak vertices from 
the  $h_{D,F,C}^{}$  terms  in  Eq.~(\ref{weak}).   
The double lines represent baryon-decuplet fields.}  
\end{figure}
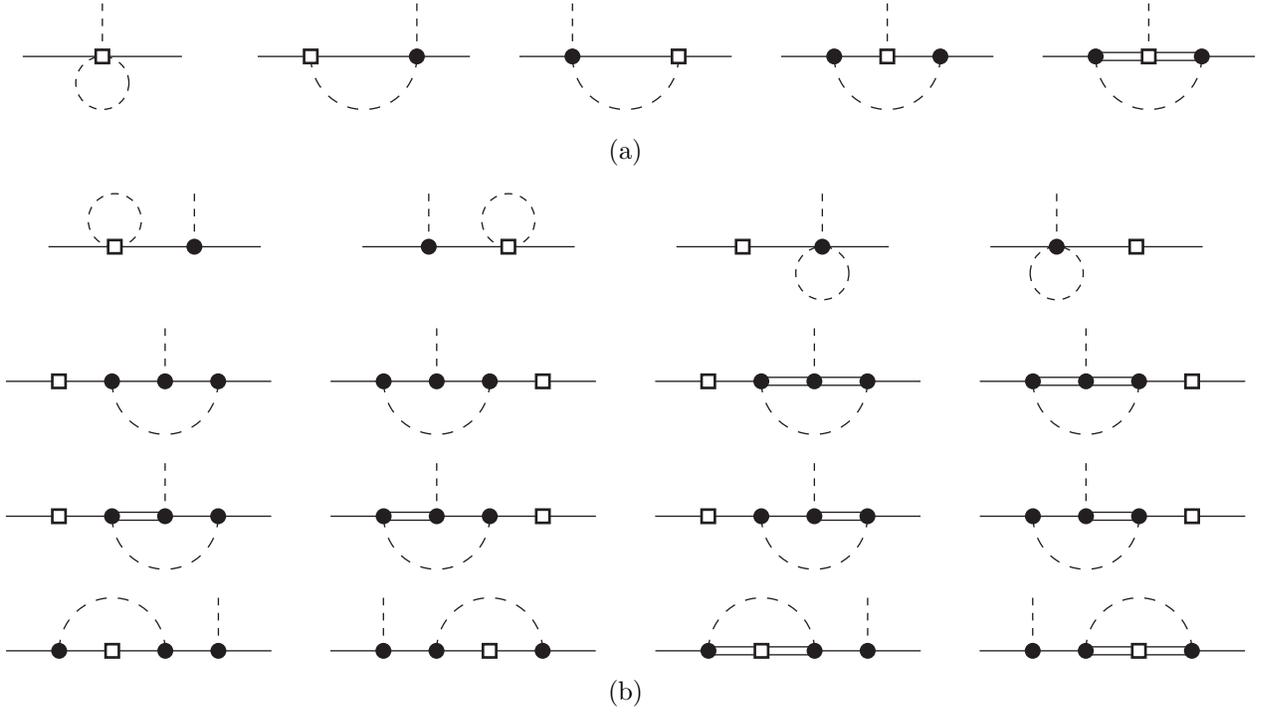             
\begin{figure}[ht]         
   \hspace*{\fill} 
\begin{picture}(80,50)(-40,0)
\Line(-30,0)(30,0) \Vertex(0,0){3}         
\DashCArc(0,10)(10,0,360){3}        
\DashLine(0,20)(0,40){3}    
\SetWidth{1}   \BBoxc(0,20)(5,5)        
\end{picture}
   \hspace*{\fill} 
\begin{picture}(80,50)(-40,0)
\Line(-40,0)(40,0) 
\DashCArc(0,0)(20,0,180){4} 
\Vertex(-20,0){3} \Vertex(20,0){3} 
\DashLine(0,20)(0,40){3}    
\SetWidth{1}   \BBoxc(0,20)(5,5) 
\end{picture}
   \hspace*{\fill} 
\begin{picture}(80,50)(-40,0)
\Line(-40,0)(-20,0) 
\Line(-20,1.5)(20,1.5) \Line(-20,-1.5)(20,-1.5) 
\Line(20,0)(40,0) 
\DashCArc(0,0)(20,0,180){4} 
\Vertex(-20,0){3} \Vertex(20,0){3} 
\DashLine(0,20)(0,40){3}    
\SetWidth{1}   \BBoxc(0,20)(5,5) 
\end{picture}
   \hspace*{\fill} 
\\   
   \hspace*{\fill} 
\begin{picture}(20,20)(-10,0)
\Text(0,5)[c]{\footnotesize(a)}   
\end{picture}
   \hspace*{\fill} 
\\   
   \hspace*{\fill} 
\begin{picture}(80,50)(-40,-20)    
\Line(-30,0)(0,0) \Vertex(0,0){3} 
\DashLine(0,0)(0,20){3} \DashCArc(0,-10)(10,90,270){3}        
\DashCArc(0,-10)(10,-90,90){3}        
\Line(0,0)(30,0)   
\SetWidth{1}   \BBoxc(0,-20)(5,5) 
\end{picture}
   \hspace*{\fill} 
\begin{picture}(80,50)(-40,-20)    
\Line(-40,0)(-20,0)   
\Line(-20,0)(0,0) \Line(0,0)(20,0) \Line(20,0)(40,0)   
\Vertex(-20,0){3} \Vertex(0,0){3} \Vertex(20,0){3}   
\DashCArc(0,0)(20,180,270){4} \DashCArc(0,0)(20,270,0){4}   
\DashLine(0,0)(0,20){3}    
\SetWidth{1}   \BBoxc(0,-20)(5,5)   
\end{picture}   
   \hspace*{\fill}   
\begin{picture}(80,50)(-40,-20)   
\Line(-40,0)(-20,0) 
\Line(-20,1.5)(0,1.5) \Line(-20,-1.5)(0,-1.5) 
\Line(0,1.5)(20,1.5)  \Line(0,-1.5)(20,-1.5)  
\Line(20,0)(40,0) 
\Vertex(-20,0){3} \Vertex(0,0){3} \Vertex(20,0){3} 
\DashCArc(0,0)(20,180,270){4} \DashCArc(0,0)(20,270,0){4} 
\DashLine(0,0)(0,20){3}   
\SetWidth{1}   \BBoxc(0,-20)(5,5) 
\end{picture}
   \hspace*{\fill} 
\begin{picture}(80,50)(-40,-20)
\Line(-40,0)(-20,0) 
\Line(-20,1.5)(0,1.5) \Line(-20,-1.5)(0,-1.5) 
\Line(0,0)(20,0)  \Line(20,0)(40,0) 
\Vertex(-20,0){3} \Vertex(0,0){3} \Vertex(20,0){3} 
\DashCArc(0,0)(20,180,270){4} \DashCArc(0,0)(20,270,0){4} 
\DashLine(0,0)(0,20){3} 
\SetWidth{1}   \BBoxc(0,-20)(5,5) 
\end{picture}
   \hspace*{\fill} 
\begin{picture}(80,50)(-40,-20)
\Line(-40,0)(-20,0) \Line(-20,0)(0,0)  
\Line(0,1.5)(20,1.5) \Line(0,-1.5)(20,-1.5) 
\Line(20,0)(40,0) 
\Vertex(-20,0){3} \Vertex(0,0){3} \Vertex(20,0){3} 
\DashCArc(0,0)(20,180,270){4} \DashCArc(0,0)(20,270,0){4} 
\DashLine(0,0)(0,20){3} 
\SetWidth{1}   \BBoxc(0,-20)(5,5) 
\end{picture}
   \hspace*{\fill} 
\\   
   \hspace*{\fill} 
\begin{picture}(20,20)(-10,0)
\Text(0,5)[c]{\footnotesize(b)}   
\end{picture}
   \hspace*{\fill} 
\caption{\label{g8,loop}%
One-loop diagrams contributing to (a) S-wave and (b) P-wave 
hyperon nonleptonic decay amplitudes, with weak vertices from   
the  $\gamma_{8}^{}$  term in  Eq.~(\ref{weak}).}  
\end{figure}
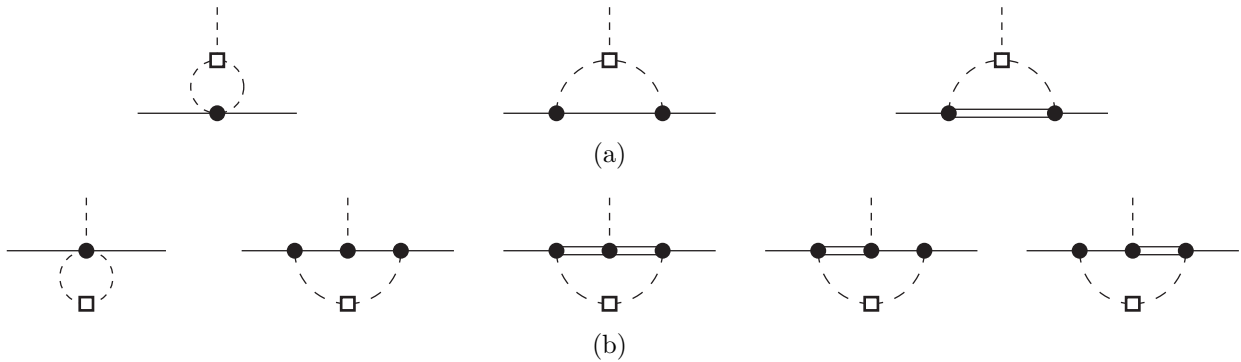             

We would now like to point out where our theoretical results differ 
from those of  Ref.~\cite{jenkins}.  
Firstly, starting from the same decay diagrams (Figures~\ref{tree},  
\ref{s-wave,p-wave,loop},  and~\ref{g8,loop})  as those used 
therein, we found the same expressions for  
$\alpha_{B_{}^{}B_{}'}^{\rm(S,P)}$,  
$\bar{\beta}_{B_{}^{}B_{}'}^{\rm(S,P)}$,  and  
$\bar{\lambda}^{}_{B_{}^{}B_{}'\pi}$,  with the exception of terms 
in  $\bar{\beta}_{B_{}^{}B_{}'}^{\rm(S,P)}$  proportional 
to~$\gamma_{8}^{}$,  corresponding to the graphs in   
Figure~\ref{g8,loop}.   
Amongst these  $\gamma_{8}^{}$~terms, we were able to reproduce 
only the expression for  $\bar{\beta}_{\Sigma^+n}^{\rm(S)}$  in  
Ref.~\cite{jenkins}.  
Secondly, we have included in the P-wave amplitudes the terms 
$\tilde{\alpha}^{\rm(P)}_{B_{}^{}B_{}'}$,  which were missing 
in Ref.~\cite{jenkins} (and in  Ref.~\cite{bsw})  
and were partially addressed in Ref.~\cite{springer}.\footnote{These 
corrections were considered for the $\,|\Delta\bfm{I}|=3/2\,$  case 
in Ref.~\cite{etv}.}  
In the next section, we will discuss how these theoretical 
modifications affect the prediction for the amplitudes.

\section{Numerical Results and Discussion\label{s4}}  
 
From the measurement of the decay rate and the decay-distribution 
asymmetry parameter $\alpha$, it is possible to extract the value 
of the S- and P-wave amplitudes for each hyperon  
decay~\cite{overseth}.  
Using the most recent data~\cite{pdb}, we find the results 
presented in Table~\ref{sx,px},\footnote{In extracting these numbers, 
final-state interactions have been ignored and experimental masses 
used.} where  $s,p$  are related to  
${\cal A}^{\rm (S,P)}_{B_{}^{}B_{}'\pi}$  by   
$\,s={\cal A}^{\rm (S)}\,$  and  
$\,p=-|\bfm{k}|{\cal A}^{\rm(P)},\,$  with  $\bfm{k}$  being 
the pion three-momentum in the rest frame of the decaying baryon.  
These numbers are nearly identical to those quoted in  
Ref.~\cite{jenkins}.

\begin{table}[ht]      
\caption{\label{sx,px}%
Experimental values for S- and P-wave amplitudes.}      
\centering   \footnotesize 
\vskip 0.5\baselineskip
\begin{tabular}{ccc}    
\hline \hline      
\raisebox{-1ex}{Decay mode $\hspace{1ex}$}   &    
\raisebox{-1ex}{$\hspace{1ex}s$}   &
\raisebox{-1ex}{$\hspace{1ex}p$}   
\vspace{1ex} \\ \hline      
\vspace{-2.5ex} & & \\  
$\begin{array}{rcl}   \displaystyle  
\Sigma^+  & \hspace{-.5em} \rightarrow & \hspace{-.5em}  n\pi^+ 
\\   
\Sigma^+  & \hspace{-.5em} \rightarrow & \hspace{-.5em}  p\pi^0   
\\      
\Sigma^-  & \hspace{-.5em} \rightarrow & \hspace{-.5em}  n\pi^-   
\\    
\Lambda   & \hspace{-.5em} \rightarrow & \hspace{-.5em}  p\pi^-    
\\        
\Lambda   & \hspace{-.5em} \rightarrow & \hspace{-.5em}  n\pi^0    
\\        
\Xi^-     & \hspace{-.5em} \rightarrow & \hspace{-.5em}  \Lambda\pi^-
\\  
\Xi^0     & \hspace{-.5em} \rightarrow & \hspace{-.5em}  \Lambda\pi^0
\end{array}$  
&  
$\begin{array}{rcl}   \displaystyle  
 0.06  & \hspace{-.5em} \pm & \hspace{-.5em}  0.01   \\  
-1.43  & \hspace{-.5em} \pm & \hspace{-.5em}  0.05   \\  
 1.88  & \hspace{-.5em} \pm & \hspace{-.5em}  0.01   \\  
 1.42  & \hspace{-.5em} \pm & \hspace{-.5em}  0.01   \\  
-1.04  & \hspace{-.5em} \pm & \hspace{-.5em}  0.02   \\  
-1.98  & \hspace{-.5em} \pm & \hspace{-.5em}  0.01   \\ 
 1.51  & \hspace{-.5em} \pm & \hspace{-.5em}  0.02   
\end{array}$  
&  
$\begin{array}{rcl}   \displaystyle   
 1.81  & \hspace{-.5em} \pm & \hspace{-.5em}  0.01   \\  
 1.17  & \hspace{-.5em} \pm & \hspace{-.5em}  0.06   \\  
-0.06  & \hspace{-.5em} \pm & \hspace{-.5em}  0.01   \\  
 0.52  & \hspace{-.5em} \pm & \hspace{-.5em}  0.01   \\  
-0.39  & \hspace{-.5em} \pm & \hspace{-.5em}  0.03   \\  
 0.48  & \hspace{-.5em} \pm & \hspace{-.5em}  0.02   \\ 
-0.32  & \hspace{-.5em} \pm & \hspace{-.5em}  0.02   
\end{array}$ 
\vspace{.3ex} \\  
\hline \hline 
\end{tabular}   
\vskip 1\baselineskip
\end{table}

To evaluate how our results describe the data, we employ 
the parameter values used in  Ref.~\cite{jenkins}:  
$\,D=0.61,$  $\,F=0.40,$  $\,{\cal C}=1.6,$  $\,{\cal H}=-1.9,$ 
$\,f_{\!P}^{}=f_{\!\pi}^{}\,,$  and  $\,\mu=1\,\rm GeV.\,$  
We also need the values of the parameters that appear 
in  Eq.~(\ref{Lmstrong}),  as they are  contained in 
$\tilde{\alpha}^{\rm(P)}_{B_{}^{}B_{}'}$.  
Following  Ref.~\cite{etv}, we choose  
\begin{eqnarray}   \label{newmasspar}     
\begin{array}{c}   \displaystyle  
b_D^{}m_s^{}  =  \ratio{3}{8} \left( m_\Sigma^{}-m_\Lambda^{} \right)  
\approx  0.0290\,{\rm GeV}   \;, 
\hspace{2em}  
b_F^{} m_s^{}  =  \ratio{1}{4} \left( m_N^{}-m_\Xi^{} \right)  
\approx  -0.0948\,{\rm GeV}   \;,  
\vspace{1ex} \\   \displaystyle   
c m_s^{}  =  \ratio{1}{2} \left( m_\Omega^{}-m_\Delta^{} \right)   
\approx  0.220\,{\rm GeV}   \;,  
\hspace{2em}  
\Delta m - 2 \left( \tilde{\sigma}-\sigma \right) m_s^{}  =  
m_\Delta^{}-m_\Sigma^{}   
\approx  0.0389\,{\rm GeV}   \;,     
\end{array}   
\end{eqnarray}    
where the fourth parameter is the only combination of 
$\Delta m$,  $\sigma m_s^{}$  and  $\tilde{\sigma} m_s^{}$   
which occurs in  $\tilde{\alpha}^{\rm(P)}_{B_{}^{}B_{}'}$.

The values of  $h_D^{}$,  $h_F^{}$,  and  $h_C^{}$  in 
Eq.~\ref{weak}  are determined by a least-squares fit using 
the four S-wave amplitudes that are not related by isospin. 
Fitting the one-loop formulas to experiment thus yields  
$\,h_D^{}=-0.84\pm0.34,\,$   $\,h_F^{}=0.78\pm0.68,\,$  and  
$\,h_C^{}=4.16\pm7.63,\,$  where all of these parameters are written 
in units of  $\,\sqrt{2}\, f_{\!\pi}^{}G_{\rm F}^{} m_{\pi^+}^2.\,$  
The quoted errors reflect an estimate of the theoretical uncertainty  
due to the terms neglected in our calculation.  
To compute them, we followed  Ref.~\cite{jenkins}  by adding 
a canonical error of  $0.30$  to each S-wave amplitude and ignoring 
the much smaller experimental error.\footnote{This is consistent with 
the fact that most of the neglected contributions are of  
${\cal O}(m_s^{})$,  which amount to corrections of about  
$\,m_s^{}/\Lambda\sim 20\%,\,$  with 
$\,\Lambda\sim 1\,\rm GeV\,$  being the chiral-symmetry breaking 
scale.}   
The uncertainty in  $h_C$  is large because it enters the amplitudes 
only at loop level and so it is poorly constrained.  
The numbers we found above are different from those found 
in Ref.~\cite{jenkins}:  
$\,h_D^{}=-0.35\pm0.09,\,$   $\,h_F^{}=0.86\pm0.05,\,$  and  
$\,h_C^{}=-0.36\pm0.65.\,$  
The discrepancy in the central values of these two sets of numbers 
indicates that the theoretical modifications we made are numerically  
important, but the large errors in the parameters, especially those 
in  $h_C^{}$  and  $h_F^{}$, make it appear less~so.  
For comparison, fitting the tree-level amplitudes gives 
$\,h_D^{}=-0.55\pm0.29\,$  and  $\,h_F^{}=1.37\pm0.17.\,$  
It is worth mentioning that, despite their variations, these three 
sets of numbers are still consistent with their expected values 
according to naive dimensional analysis~\cite{nda}.

Using the parameter values from our fit above, we obtained 
the numerical results presented in  Table~\ref{spx,spt,ind}.  
Since the S-wave formula for  $\,\Sigma^+\rightarrow n\pi^+\,$  
does not depend on  $h_{D,F,C}^{}$,  the three-parameter fit leads 
to exact agreement with experiment for the other three S-wave 
channels.  
As a consequence, the Lee-Sugawara relation~\cite{LeeSug},  
$\,3s^{}_{\Sigma^-\rightarrow n\pi^-}/\sqrt{6}
+ s^{}_{\Lambda\rightarrow p\pi^-}  
+ 2 s^{}_{\Xi^-\rightarrow \Lambda\pi^-}=0,\,$  
which is a prediction of SU(3) symmetry and agrees well with data,    
is well satisfied.   
A good fit (using all of the seven S-waves) was also obtained in  
Ref.~\cite{jenkins}, but the individual tree and loop contributions 
therein are numerically different from ours, in some cases markedly, 
albeit within expectations.      
As in  Ref.~\cite{jenkins},  we can see in  Table~\ref{spx,spt,ind}   
that some of the loop corrections in the S-waves are comparable in 
size to the lowest-order results even though they are naively   
expected to be smaller by about a factor of  $20\%$.   
In addition, we found that  ${\cal O}(m_s^2\ln m_s^{})$  
contributions (from diagrams proportional to~$\gamma_8^{}$)  are 
sometimes larger than those of  ${\cal O}(m_s^{}\ln m_s^{})$.  
This lack of convergence is an inherent flaw in a perturbative 
calculation where the expansion parameter,  $m_s^{}/\Lambda$,  is 
not sufficiently small and there are many loop-diagrams involved. 
The problem also occurs in the  $\,|\Delta\bfm{I}|=3/2\,$   
sector~\cite{etv}  and in other cases~\cite{problem}. 
For the P-waves, the disagreement between theory and experiment is 
worse than before.  
The tree-level contributions remain suppressed with respect to  
the chiral-logarithmic corrections because of the near cancellations  
of the two terms in the tree-level formulas~\cite{jenkins},
and the loop correction receives a sizable contribution from 
the new term  $\tilde{\alpha}^{\rm(P)}_{B_{}^{}B_{}'}$.

\begin{table}[ht]      
\caption{\label{spx,spt,ind}%
Experimental and theoretical values of S- and P-wave amplitudes 
for  $\,h_D^{}=-0.84,\,$  $\,h_F^{}=0.78,\,$  and  
$\,h_C^{}=4.16.\,$}
\centering   \footnotesize 
\vskip 0.5\baselineskip   
\begin{tabular}{ccccccc}    
\hline \hline      
\raisebox{-1ex}{Decay mode $\hspace{1ex}$}   &    
\raisebox{-1ex}{$\hspace{1ex}s_{\rm expt}^{}$}   &   
\raisebox{-1ex}{$\hspace{1ex}s_{\rm theory}^{}$}   &
\raisebox{-1ex}{$\hspace{1ex}s_{\rm tree}^{}$}   &
\raisebox{-1ex}{$\hspace{1ex}s_{\rm loop}^{}$}  & 
\raisebox{-1ex}{$\hspace{1ex}s_{\rm loop}^{\rm(oct)}$}   &
\raisebox{-1ex}{$\hspace{1ex}s_{\rm loop}^{\rm(dec)}$}   
\vspace{0.5ex} \\ \hline      
\vspace{-2.5ex} \\  
$\begin{array}{rcl}   \displaystyle  
\Sigma^+  & \hspace{-.5em} \rightarrow & \hspace{-.5em}  n\pi^+   
\\   
\Sigma^-  & \hspace{-.5em} \rightarrow & \hspace{-.5em}  n\pi^-   
\\    
\Lambda   & \hspace{-.5em} \rightarrow & \hspace{-.5em}  p\pi^-    
\\        
\Xi^-     & \hspace{-.5em} \rightarrow & \hspace{-.5em}  \Lambda\pi^-  
\end{array}$  
&  
$\begin{array}{r}   \displaystyle  
 0.06 \\  1.88 \\  1.42 \\  -1.98   
\end{array}$  
&  
$\begin{array}{r}   \displaystyle  
-0.09 \\  1.88 \\  1.42  \\ -1.98  
\end{array}$  
&  
$\begin{array}{r}   \displaystyle  
 0.00 \\  1.62 \\  0.61 \\ -1.29  
\end{array}$  
&  
$\begin{array}{r}   \displaystyle  
-0.09 \\  0.26 \\  0.81 \\ -0.69 
\end{array}$  
&  
$\begin{array}{r}   \displaystyle  
 0.13 \\  0.40 \\  0.24 \\ -0.22  
\end{array}$  
&  
$\begin{array}{r}   \displaystyle  
-0.22 \\ -0.14 \\  0.58 \\ -0.46 
\end{array}$ 
\vspace{.3ex} \\  
\hline \hline 
\raisebox{-1ex}{Decay mode $\hspace{1ex}$}   &    
\raisebox{-1ex}{$\hspace{1ex}p_{\rm expt}^{}$}   &   
\raisebox{-1ex}{$\hspace{1ex}p_{\rm theory}^{}$}   &
\raisebox{-1ex}{$\hspace{1ex}p_{\rm tree}^{}$}   &
\raisebox{-1ex}{$\hspace{1ex}p_{\rm loop}^{}$}  & 
\raisebox{-1ex}{$\hspace{1ex}p_{\rm loop}^{\rm(oct)}$}   &
\raisebox{-1ex}{$\hspace{1ex}p_{\rm loop}^{\rm(dec)}$}   
\vspace{0.5ex} \\ \hline      
\vspace{-2.5ex} \\  
$\begin{array}{rcl}   \displaystyle  
\Sigma^+  & \hspace{-.5em} \rightarrow & \hspace{-.5em}  n\pi^+   
\\   
\Sigma^-  & \hspace{-.5em} \rightarrow & \hspace{-.5em}  n\pi^-   
\\    
\Lambda   & \hspace{-.5em} \rightarrow & \hspace{-.5em}  p\pi^-    
\\        
\Xi^-     & \hspace{-.5em} \rightarrow & \hspace{-.5em}  \Lambda\pi^-  
\end{array}$  
&  
$\begin{array}{r}   \displaystyle  
 1.81 \\ -0.06 \\  0.52 \\  0.48 
\end{array}$  
&  
$\begin{array}{r}   \displaystyle  
 2.41 \\  1.93 \\ -1.13 \\  2.17  
\end{array}$  
&  
$\begin{array}{r}   \displaystyle  
-0.40 \\ -0.16 \\ -0.03 \\ -0.22 
\end{array}$  
&  
$\begin{array}{r}   \displaystyle  
 2.81 \\  2.10 \\ -1.10 \\  2.39  
\end{array}$  
&  
$\begin{array}{r}   \displaystyle  
 0.07 \\  0.08 \\ -0.09 \\  0.06  
\end{array}$  
&  
$\begin{array}{r}   \displaystyle  
 2.74 \\  2.01 \\ -1.01 \\  2.33  
\end{array}$ 
\vspace{.3ex} \\  
\hline \hline 
\end{tabular}   
\end{table}

The dependence of the one-loop contributions in  Eq.~(\ref{As,Ap}) 
on the renormalization scale  $\mu$  is canceled by 
the $\mu$-dependence of the counterterms that have been neglected, 
and so our one-loop amplitudes depend on the choice of  $\mu$.  
It is, therefore, important to know how our results change as  $\mu$ 
varies.  
At the same time, we would also like to know how our results will 
differ if we make the following two changes in order to reduce, at  
least at one-loop order, the effects of kaon and eta loops, which  
may be overestimated in  $\chi$PT~\cite{jlms}.  
We will set  $\,f_{\!P}^{}=f_K^{}\approx 113\,\rm MeV,\,$  instead 
of  $f_{\!\pi}^{}$,  in the amplitudes in  Eq.~(\ref{As,Ap}), 
for the difference will appear at higher orders.  
Also, we will use the same values of  $D$, $F$,  and  
${\cal H}$  as before, but now  ${\cal C}=1.2,\,$  all of which 
are the favored values from one-loop fits to semileptonic and  
strong hyperon decays~\cite{JenMan3,bss}.   
We will consider three different values of  $\mu$, and, 
for each of them, perform a least-squares fit as before 
to determine the values of  $h_{D,F,C}^{}$.   
The results of these fits,  along with the corresponding predictions 
for the amplitudes, are given in  Table~\ref{spx,spt,fk,ind}.

\begin{table}[ht]    
\caption{\label{spx,spt,fk,ind}%
Experimental and theoretical values of S- and P-wave amplitudes for  
various values of $\mu$,  and the corresponding values of  $h_D^{}$,  
$h_F^{}$,  and  $h_C^{}$.}     
\centering   \footnotesize 
\vskip 0.5\baselineskip   
\begin{tabular}{c|cccc|cccc}    
\hline \hline      
\multicolumn{9}{c}{\raisebox{-0.5ex}{$\,\mu=0.8\,\rm GeV,\,$  
$\,h_D^{}=-0.76\pm0.37,\,$  $\,h_F^{}=0.89\pm0.74,\,$  
$\,h_C^{}=8.6\pm20.4\,$}}  
\vspace{0.5ex} \\ \hline      
\raisebox{-0.5ex}{Decay mode $\hspace{1ex}$}       &    
\raisebox{-0.5ex}{$\hspace{1ex}s_{\rm expt}^{}$}   &   
\raisebox{-0.5ex}{$\hspace{1ex}s_{\rm theory}^{}$} &  
\raisebox{-0.5ex}{$\hspace{1ex}s_{\rm tree}^{}$}   &  
\raisebox{-0.5ex}{$\hspace{1ex}s_{\rm loop}^{}$}   &  
\raisebox{-0.5ex}{$\hspace{1ex}p_{\rm expt}^{}$}   &   
\raisebox{-0.5ex}{$\hspace{1ex}p_{\rm theory}^{}$} &  
\raisebox{-0.5ex}{$\hspace{1ex}p_{\rm tree}^{}$}   &  
\raisebox{-0.5ex}{$\hspace{1ex}p_{\rm loop}^{}$}   
\vspace{0.5ex} \\ \hline      
\vspace{-2.5ex} \\  
$\begin{array}{rcl}   \displaystyle  
\Sigma^+  & \hspace{-.5em} \rightarrow & \hspace{-.5em}  n\pi^+   
\\      
\Sigma^-  & \hspace{-.5em} \rightarrow & \hspace{-.5em}  n\pi^-   
\\    
\Lambda   & \hspace{-.5em} \rightarrow & \hspace{-.5em}  p\pi^-    
\\        
\Xi^-     & \hspace{-.5em} \rightarrow & \hspace{-.5em}  \Lambda\pi^-  
\end{array}$  
&  
$\begin{array}{r}   \displaystyle  
 0.06 \\  1.88 \\  1.42 \\ -1.98 
\end{array}$  
&  
$\begin{array}{r}   \displaystyle  
 0.00 \\  1.88 \\  1.42 \\ -1.98 
\end{array}$  
&  
$\begin{array}{r}   \displaystyle  
 0.00 \\  1.65 \\  0.78 \\ -1.40 
\end{array}$  
&  
$\begin{array}{r}   \displaystyle  
 0.00 \\  0.23 \\  0.64 \\ -0.58 
\end{array}$  
&  
$\begin{array}{r}   \displaystyle  
 1.81 \\ -0.06 \\  0.52 \\  0.48 
\end{array}$  
&  
$\begin{array}{r}   \displaystyle  
 0.76 \\  0.69 \\ -0.55 \\  1.06 
\end{array}$  
&  
$\begin{array}{r}   \displaystyle  
-0.33 \\ -0.08 \\ -0.12 \\ -0.13 
\end{array}$  
&  
$\begin{array}{r}   \displaystyle  
 1.09 \\  0.77 \\ -0.43 \\  1.19 
\end{array}$  
\vspace{0.3ex} \\  
\hline \hline 
\end{tabular}   
\begin{tabular}{c|cccc|cccc}    
\multicolumn{9}{c}{\raisebox{-0.5ex}{$\,\mu=1\,\rm GeV,\,$  
$\,h_D^{}=-0.78\pm0.33,\,$  $\,h_F^{}=0.82\pm0.68,\,$  
$\,h_C^{}=7.2\pm14.7.\,$}}    
\vspace{0.5ex} \\ \hline      
\raisebox{-0.5ex}{Decay mode $\hspace{1ex}$}       &    
\raisebox{-0.5ex}{$\hspace{1ex}s_{\rm expt}^{}$}   &   
\raisebox{-0.5ex}{$\hspace{1ex}s_{\rm theory}^{}$} &
\raisebox{-0.5ex}{$\hspace{1ex}s_{\rm tree}^{}$}   &
\raisebox{-0.5ex}{$\hspace{1ex}s_{\rm loop}^{}$}   & 
\raisebox{-0.5ex}{$\hspace{1ex}p_{\rm expt}^{}$}   &   
\raisebox{-0.5ex}{$\hspace{1ex}p_{\rm theory}^{}$} &
\raisebox{-0.5ex}{$\hspace{1ex}p_{\rm tree}^{}$}   &
\raisebox{-0.5ex}{$\hspace{1ex}p_{\rm loop}^{}$}  
\vspace{0.5ex} \\ \hline      
\vspace{-2.5ex} \\  
$\begin{array}{rcl}   \displaystyle  
\Sigma^+  & \hspace{-.5em} \rightarrow & \hspace{-.5em}  n\pi^+   
\\      
\Sigma^-  & \hspace{-.5em} \rightarrow & \hspace{-.5em}  n\pi^-   
\\    
\Lambda   & \hspace{-.5em} \rightarrow & \hspace{-.5em}  p\pi^-    
\\        
\Xi^-     & \hspace{-.5em} \rightarrow & \hspace{-.5em}  \Lambda\pi^-  
\end{array}$  
&  
$\begin{array}{r}   \displaystyle  
 0.06 \\  1.88 \\  1.42 \\ -1.98 
\end{array}$  
&  
$\begin{array}{r}   \displaystyle  
 0.00 \\  1.88 \\  1.42 \\ -1.98 
\end{array}$  
&  
$\begin{array}{r}   \displaystyle  
 0.00 \\  1.60 \\  0.68 \\ -1.32 
\end{array}$  
&  
$\begin{array}{r}   \displaystyle  
 0.00 \\  0.28 \\  0.74 \\ -0.66 
\end{array}$  
&  
$\begin{array}{r}   \displaystyle  
 1.81 \\ -0.06 \\  0.52 \\  0.48 
\end{array}$  
&  
$\begin{array}{r}   \displaystyle  
 0.82 \\  0.80 \\ -0.65 \\  1.21 
\end{array}$  
&  
$\begin{array}{r}   \displaystyle  
-0.36 \\ -0.12 \\ -0.08 \\ -0.17 
\end{array}$  
&  
$\begin{array}{r}   \displaystyle  
 1.17 \\  0.91 \\ -0.58 \\  1.38 
\end{array}$  
\vspace{.3ex} \\  
\hline \hline 
\end{tabular}   
\begin{tabular}{c|cccc|cccc}    
\multicolumn{9}{c}{\raisebox{-0.5ex}{$\,\mu=1.2\,\rm GeV,\,$  
$\,h_D^{}=-0.80\pm0.30,\,$  $\,h_F^{}=0.77\pm0.64,\,$  
$\,h_C^{}=6.7\pm12.1.\,$}}    
\vspace{0.5ex} \\ \hline      
\raisebox{-0.5ex}{Decay mode $\hspace{1ex}$}       &    
\raisebox{-0.5ex}{$\hspace{1ex}s_{\rm expt}^{}$}   &   
\raisebox{-0.5ex}{$\hspace{1ex}s_{\rm theory}^{}$} &
\raisebox{-0.5ex}{$\hspace{1ex}s_{\rm tree}^{}$}   &
\raisebox{-0.5ex}{$\hspace{1ex}s_{\rm loop}^{}$}   & 
\raisebox{-0.5ex}{$\hspace{1ex}p_{\rm expt}^{}$}   &   
\raisebox{-0.5ex}{$\hspace{1ex}p_{\rm theory}^{}$} &
\raisebox{-0.5ex}{$\hspace{1ex}p_{\rm tree}^{}$}   &
\raisebox{-0.5ex}{$\hspace{1ex}p_{\rm loop}^{}$}   
\vspace{0.5ex} \\ \hline      
\vspace{-2.5ex} \\  
$\begin{array}{rcl}   \displaystyle  
\Sigma^+  & \hspace{-.5em} \rightarrow & \hspace{-.5em}  n\pi^+   
\\   
\Sigma^-  & \hspace{-.5em} \rightarrow & \hspace{-.5em}  n\pi^-   
\\    
\Lambda   & \hspace{-.5em} \rightarrow & \hspace{-.5em}  p\pi^-    
\\        
\Xi^-     & \hspace{-.5em} \rightarrow & \hspace{-.5em}  \Lambda\pi^-  
\end{array}$  
&  
$\begin{array}{r}   \displaystyle  
 0.06 \\  1.88 \\  1.42 \\ -1.98 
\end{array}$  
&  
$\begin{array}{r}   \displaystyle  
 0.00 \\  1.88 \\  1.42 \\ -1.98 
\end{array}$  
&  
$\begin{array}{r}   \displaystyle  
 0.00 \\  1.57 \\  0.62 \\ -1.27 
\end{array}$  
&  
$\begin{array}{r}   \displaystyle  
 0.00 \\  0.31 \\  0.80 \\ -0.71 
\end{array}$  
&  
$\begin{array}{r}   \displaystyle  
 1.81 \\ -0.06 \\  0.52 \\  0.48 
\end{array}$  
&  
$\begin{array}{r}   \displaystyle  
 0.86 \\  0.88 \\ -0.72 \\  1.34 
\end{array}$  
&  
$\begin{array}{r}   \displaystyle  
-0.38 \\ -0.14 \\ -0.05 \\ -0.20 
\end{array}$  
&  
$\begin{array}{r}   \displaystyle  
 1.24 \\  1.02 \\ -0.68 \\  1.53 
\end{array}$  
\vspace{.3ex} \\  
\hline \hline 
\end{tabular}   
\end{table}

We can see that the central values of  $h_{D,F}^{}$  are relatively 
stable with respect to changes in the other parameters, and  
$h_{C}^{}$  is less so, but the different values of these weak 
parameters are still consistent with each other in view of 
their large errors.   
The tree and loop terms of the S-waves are also relatively stable 
against the parameter changes.  
The loop terms of the P-waves, however, change significantly as we 
move from  Table~\ref{spx,spt,ind}  to  Table~\ref{spx,spt,fk,ind}, 
whereas the P-waves in  Table~\ref{spx,spt,fk,ind}  do not as  
$\mu$  is varied.  
This significant change is mainly due to our choice of  
$\,f_{\!P}^{}=f_K^{}\,$  and  $\,{\cal C}=1.2\,$  for  
Table~\ref{spx,spt,fk,ind}, which leads to a dramatic decrease of 
the loop contributions with respect to the leading-order terms,  
alleviating the discrepancy between theory and experiment.   
Comparison of the  $\,\mu=1\,\rm GeV\,$  cases in the two tables 
shows that this choice also leads to a slight reduction in the lack of  
chiral convergence in our S-wave formulas.  
Finally, we should mention that our one-loop formulas, with the sets 
of parameter values used in  Table~\ref{spx,spt,fk,ind}, can describe 
both the (seven) S- and P-wave data better than either a tree-level 
fit or the one-loop fit of Ref.~\cite{jenkins}  can, although 
the P-waves remain poorly reproduced.\footnote{%
A recent study~\cite{BorHol'} on the role of baryon resonances 
in these decays has suggested that including counterterms is   
important for a satisfactory description of both the S- and P-waves 
in chiral perturbation theory.}

In conclusion, we have reexamined the one-loop analysis of 
the amplitudes for hyperon nonleptonic decays in chiral perturbation 
theory, concentrating on the leading nonanalytic contributions to 
the amplitudes. 
We have discussed how our theoretical results differed from those 
previously calculated using the same approach. 
Even though the differences are numerically important, they do not 
alter the well-known fact that a good prediction at one-loop level 
cannot simultaneously be made for the S- and P-wave amplitudes.  
Nevertheless, our results suggest that a judicious choice of 
the parameter values in the theory can, at least at one-loop level, 
lead to a sizable reduction of the large kaon-loop effects relative   
to the lowest-order contributions and yield an improved fit to   
experiment.

{\it Note added}$\;$  
After submitting this paper for publication, we became aware of 
Ref.~\cite{springer'}, in which one-loop corrections 
to the propagator in the tree-level P-waves are also considered 
and added to the amplitudes calculated in  Ref.~\cite{jenkins}.  
The expressions obtained in  Ref.~\cite{springer'}  for the new   
contributions disagree with ours, but we have agreement in that 
the P-waves remain poorly reproduced.

\bigskip

{\bf Acknowledgments}$\;$    
We would like to thank G.~Valencia for many helpful discussions and 
suggestions. 
J.~T. thanks the Fermilab Theory Group for hospitality during its 
Summer Visitors Program while part of this work was done.  
This work was supported in part by DOE under contract number 
DE-FG02-92ER40730. 
The work of A.~A. was supported in part by DOE under contract 
number DE-FG02-87ER40371, and by the International Institute of Theoretical
and Applied Physics, Iowa State University, Ames, Iowa.

\newpage

\noindent
{\Large\bf Appendix}  

From the tree-level diagrams displayed in Figure~\ref{tree}, 
\begin{eqnarray*}        
\alpha^{(\rm S)}_{\Sigma^+ n}  =  0   
\;, \hspace{1em}         
\alpha^{(\rm S)}_{\Sigma^- n}  =  -h_D^{} + h_F^{}   
\;, \hspace{1em}         
\alpha^{(\rm S)}_{\Lambda p}  \;=\;  
\ratio{1}{\sqrt{6}} \bigl( h_D^{}+3 h_F^{} \bigr)    
\;, \hspace{1em}         
\alpha^{(\rm S)}_{\Xi^-\Lambda}  \;=\;  
\ratio{1}{\sqrt{6}} \bigl( h_D^{}-3 h_F^{} \bigr)    \;,           
\end{eqnarray*}    
\begin{eqnarray*}        
\begin{array}{c}   \displaystyle  
\alpha^{(\rm P)}_{\Sigma^+ n}  \;=\;   \displaystyle  
{-D\, \bigl( h_D^{}-h_F^{} \bigr) \over m_\Sigma^{}-m_N^{}}   
- { \ratio{1}{3} D\, \bigl( h_D^{}+3 h_F^{} \bigr) 
   \over  m_\Lambda^{}-m_N^{} }   \;, 
\vspace{1ex} \\   \displaystyle    
\alpha^{(\rm P)}_{\Sigma^- n}  \;=\;   \displaystyle  
{-F\, \bigl( h_D^{}-h_F^{} \bigr) \over m_\Sigma^{}-m_N^{}}   
- { \ratio{1}{3} D\, \bigl( h_D^{}+3 h_F^{} \bigr) 
   \over  m_\Lambda^{}-m_N^{} }   \;,     
\vspace{1ex} \\   \displaystyle    
\alpha^{(\rm P)}_{\Lambda p}     \;=\;   \displaystyle      
{ 2 D\, \bigl( h_D^{}-h_F^{} \bigr)  
 \over  \sqrt{6}\; \bigl( m_\Sigma^{}-m_N^{} \bigr) }   
+ { (D+F)\, \bigl( h_D^{}+3 h_F^{} \bigr)  
   \over  \sqrt{6}\; \bigl( m_\Lambda^{}-m_N^{} \bigr) }   \;, 
\vspace{1ex} \\   \displaystyle    
\alpha^{(\rm P)}_{\Xi^-\Lambda}  \;=\;   \displaystyle   
{ -2 D\, \bigl( h_D^{}+h_F^{} \bigr)  
  \over  \sqrt{6}\; \bigl( m_\Xi^{}-m_\Sigma^{} \bigr) }   
- { (D-F)\, \bigl( h_D^{}-3 h_F^{} \bigr) 
   \over  \sqrt{6}\; \bigl( m_\Xi^{}-m_\Lambda^{} \bigr) }   \;. 
\end{array}    
\end{eqnarray*}    
From one-loop diagrams involving only octet baryons, shown in 
Figures~\ref{s-wave,p-wave,loop} and~\ref{g8,loop}, 
\begin{eqnarray*}   \label{octetsl}  
\begin{array}{rl}   \displaystyle  
& 
\beta^{(\rm S)}_{\Sigma^+ n}  \;=\;     
-2 D^2\, (m_\Sigma^{}-m_N^{}) \gamma_{8}^{}   \;,        
\vspace{2ex} \\   \displaystyle   
\beta^{(\rm S)}_{\Sigma^- n}   \;=&    
\ratio{11}{12} \bigl( h_D^{}-h_F^{} \bigr) 
+ \Bigl( \ratio{7}{6} D^2-3 D F-\ratio{3}{2} F^2 \Bigr) h_D^{}  
+ \Bigl( \ratio{5}{6} D^2+5 D F+\ratio{3}{2} F^2 \Bigr) h_F^{}   
\vspace{1ex} \nonumber \\  & 
+\;   
\Bigl[ \left( \ratio{7}{6} + 4 D^2 - 9 D F + 3 F^2 \right) 
      (m_\Sigma^{}-m_N^{})  
      + \left( D^2 + 3 D F \right) (m_\Lambda^{}-m_N^{}) \Bigr]    
\gamma_{8}^{}   \;,  
\vspace{2ex} \\   \displaystyle   
\beta^{(\rm S)}_{\Lambda p}     \;=&    
\ratio{1}{\sqrt{6}} \Bigl[ 
-\ratio{11}{12} \bigl( h_D^{}+3 h_F^{} \bigr) 
+ \Bigl( \ratio{19}{6} D^2-11 D F+\ratio{9}{2} F^2 \Bigr) h_D^{}  
+ \Bigl( \ratio{7}{2} D^2-15 D F+\ratio{27}{2} F^2 \Bigr) h_F^{}   
\Bigr]   
\vspace{1ex} \nonumber \\ & 
+\;   
\ratio{1}{\sqrt{6}} 
\Bigl[ \left( \ratio{7}{2} + 2 D^2 - 3 D F + 9 F^2 \right) 
      \bigl( m_\Lambda^{}-m_N^{} \bigr)   
      + \left( -9 D^2 + 9 D F \right) (m_\Sigma^{}-m_N^{})  
\Bigr] \gamma_{8}^{}   \;,    
\vspace{2ex} \\   \displaystyle   
\beta^{(\rm S)}_{\Xi^-\Lambda}  \;=&    
\ratio{1}{\sqrt{6}} \Bigl[ 
-\ratio{11}{12} \bigl( h_D^{}-3 h_F^{} \bigr) 
+ \Bigl( \ratio{19}{6} D^2+11 D F+\ratio{9}{2} F^2 \Bigr) h_D^{}  
- \Bigl( \ratio{7}{2} D^2+15 D F+\ratio{27}{2} F^2 \Bigr) h_F^{}   
\Bigr]   
\vspace{1ex} \nonumber \\ & 
+\;   
\ratio{1}{\sqrt{6}} 
\Bigl[ \left( -\ratio{7}{2} + 7 D^2 + 6 D F - 9 F^2 \right) 
      \bigl( m_\Xi^{}-m_\Lambda^{} \bigr)   
      + \left( -9 D^2 - 9 D F \right) (m_\Sigma^{}-m_\Lambda^{})  
\Bigr] \gamma_{8}^{}   \;,    
\end{array}
\end{eqnarray*}

\newpage

\begin{eqnarray*}   \label{octetpl1}  
\beta^{(\rm P)}_{\Sigma^+ n}  &\!\!=&\!\!  
{ \ratio{17}{12} D\, \bigl( h_D^{}-h_F^{} \bigr)  \over  
 m_\Sigma^{}-m_N^{}}   
+ { \ratio{17}{36} D\, \bigl( h_D^{}+3 h_F^{} \bigr)  \over  
   m_\Lambda^{}-m_N^{}}   
\nonumber \\ &&   
+\;   
{ \Bigl( \ratio{17}{18} D^3 - \ratio{19}{6} D^2 F 
        - \ratio{13}{6} D F^2 + \ratio{3}{2} F^3 \Bigr) h_D^{}  \over   
 m_\Sigma^{}-m_N^{} } 
+ { \Bigl( \ratio{19}{18} D^3+\ratio{31}{6} D^2 F+\ratio{13}{6} D F^2 
          - \ratio{3}{2} F^3 \Bigr) h_F^{}  \over  m_\Sigma^{}-m_N^{} } 
\nonumber \\ && 
+\;   
{ \Bigl( -\ratio{37}{27} D^3 + \ratio{11}{3} D^2 F 
        - \ratio{4}{3} D F^2 \Bigr) h_D^{}  \over   
 m_\Lambda^{}-m_N^{} } 
- { \Bigl( \ratio{19}{9} D^3-5 D^2 F+4 D F^2 \Bigr) h_F^{}  \over  
   m_\Lambda^{}-m_N^{} }   
\nonumber \\ && 
+\;   
\Bigl( \ratio{49}{27} D^3 - D^2 F - \ratio{19}{9} D F^2 
      + \ratio{11}{3} F^3 \Bigr) \gamma_{8}^{}   \;,      
\end{eqnarray*}
\begin{eqnarray*}   \label{octetpl2}  
\beta^{(\rm P)}_{\Sigma^- n}  &\!\!=&\!\!  
{ \ratio{17}{12} F\, \bigl( h_D^{}-h_F^{} \bigr)  \over  
 m_\Sigma^{}-m_N^{} } 
+ { \ratio{17}{36} D\, \bigl( h_D^{}+3 h_F^{} \bigr)  \over   
   m_\Lambda^{}-m_N^{} } 
\nonumber \\ && 
+\;   
{ \Bigl( \ratio{10}{9} D^2 F-3 D F^2 - F^3 \Bigr) h_D^{} 
 \over  m_\Sigma^{}-m_N^{} } 
+ { \Bigl( \ratio{8}{9} D^2 F+5 D F^2 + F^3 \Bigr) h_F^{} 
   \over  m_\Sigma^{}-m_N^{} } 
\nonumber \\ && 
+\;   
{ \Bigl( -\ratio{37}{27} D^3 + \ratio{11}{3} D^2 F - 
        \ratio{4}{3} D F^2 \Bigr) h_D^{}  \over  
 m_\Lambda^{}-m_N^{} } 
- { \Bigl( \ratio{19}{9} D^3 - 5 D^2 F + 4 D F^2 \Bigr) h_F^{} \over  
   m_\Lambda^{}-m_N^{} }   
\nonumber \\ && 
+\;   
\Bigl( \ratio{11}{18} D - \ratio{11}{18} F + \ratio{34}{27} D^3 
      + \ratio{20}{9} D^2 F - \ratio{34}{9} D F^2 \Bigr)   
\gamma_{8}^{}   \;,      
\\ \nonumber \\   
\label{octetpl3}  
\beta^{(\rm P)}_{\Lambda p}  &\!\!=&\!\!  
-{ \ratio{17}{6} D\, \bigl( h_D^{}-h_F^{} \bigr)  \over  
  \sqrt{6}\; \bigl( m_\Sigma^{}-m_N^{} \bigr) } 
- { \ratio{17}{12} (D+F)\, \bigl( h_D^{}+3 h_F^{} \bigr)  \over  
   \sqrt{6}\; \bigl( m_\Lambda^{}-m_N^{} \bigr) } 
\nonumber \\ && 
+\;   
{ \Bigl( -\ratio{4}{9} D^3 + 6 D^2 F + 2 D F^2 \Bigr) h_D^{}  \over  
 \sqrt{6}\; \bigl( m_\Sigma^{}-m_N^{} \bigr) } 
- { \Bigl( \ratio{32}{9} D^3 + 10 D^2 F + 2 D F^2 \Bigr) h_F^{}  \over  
   \sqrt{6}\; \bigl( m_\Sigma^{}-m_N^{} \bigr) } 
\nonumber \\ && 
+\;   
{ \Bigl( \ratio{61}{18} D^3 - \ratio{137}{18} D^2 F
        - \ratio{35}{6} D F^2 + \ratio{5}{2} F^3 \Bigr) h_D^{}  
 \over  \sqrt{6}\; \bigl( m_\Lambda^{}-m_N^{} \bigr) } 
+ { \Bigl( \ratio{25}{6} D^3 - \ratio{65}{6} D^2 F
          + \ratio{1}{2} D F^2 + \ratio{15}{2} F^3 \Bigr) h_F^{}  
   \over  \sqrt{6}\; \bigl( m_\Lambda^{}-m_N^{} \bigr) }   
\nonumber \\ && 
+\;   
\ratio{1}{\sqrt{6}} \Bigl( 
-\ratio{11}{18} D - \ratio{11}{6} F   
+ D^3 + \ratio{23}{9} D^2 F + 3 D F^2 + 5 F^3   
\Bigr) \gamma_{8}^{}   \;,      
\\ \nonumber \\   
\label{octetpl4}        
\beta^{(\rm P)}_{\Xi^-\Lambda}  &\!\!=&\!\!  
{ \ratio{17}{6} D\, \bigl( h_D^{}+h_F^{} \bigr)  
 \over  \sqrt{6}\; \bigl( m_\Xi^{}-m_\Sigma^{} \bigr) } 
+ { \ratio{17}{12} (D-F)\, \bigl( h_D^{}-3 h_F^{} \bigr) 
   \over  \sqrt{6}\; \bigl( m_\Xi^{}-m_\Lambda^{} \bigr) }   
\nonumber \\ && 
+\;   
{ \Bigl( \ratio{4}{9} D^3+6 D^2 F-2 D F^2 \Bigr) h_D^{}  
 \over  \sqrt{6}\; \bigl( m_\Xi^{}-m_\Sigma^{} \bigr) } 
- { \Bigl( \ratio{32}{9} D^3-10 D^2 F+2 D F^2 \Bigr) h_F^{}  
   \over  \sqrt{6}\; \bigl( m_\Xi^{}-m_\Sigma^{} \bigr) } 
\nonumber \\ && 
-\;   
{ \Bigl( \ratio{61}{18} D^3+\ratio{137}{18} D^2 F
        - \ratio{35}{6} D F^2-\ratio{5}{2} F^3 \Bigr) h_D^{}  
 \over  \sqrt{6}\; \bigl( m_\Xi^{}-m_\Lambda^{} \bigr) } 
+ { \Bigl( \ratio{25}{6} D^3+\ratio{65}{6} D^2 F
          + \ratio{1}{2} D F^2-\ratio{15}{2} F^3 \Bigr) h_F^{}  
   \over  \sqrt{6}\; \bigl( m_\Xi^{}-m_\Lambda^{} \bigr) }   
\nonumber \\ && 
+\;   
\ratio{1}{\sqrt{6}} \Bigl( 
-\ratio{11}{18} D + \ratio{11}{6} F   
+ D^3 - \ratio{23}{9} D^2 F + 3 D F^2 - 5 F^3 
\Bigr) \gamma_{8}^{}   \;. 
\end{eqnarray*}
From one-loop diagrams involving decuplet baryons, also shown in 
Figures~\ref{s-wave,p-wave,loop} and~\ref{g8,loop}, 
\begin{eqnarray*}         
\begin{array}{rl}   \displaystyle  
& 
\beta^{\prime(\rm S)}_{\Sigma^+ n}  \;=\;  
\ratio{1}{2}\, {\cal C}^2\, (m_\Sigma^{}-m_N^{})\, \gamma_{8}^{}   \;,  
\vspace{1ex} \\   \displaystyle   
\beta^{\prime(\rm S)}_{\Sigma^- n}  \;=&   
-\ratio{1}{9}\, {\cal C}^2\, h_C^{}   
\;+\;  
{\cal C}^2\, 
\Bigl[ -\ratio{23}{18} (m_\Sigma^{}-m_N^{}) 
      + \ratio{8}{3} (m_\Delta^{}-m_N^{})  
      + \ratio{13}{9} \bigl( m_{\Sigma^*}^{}-m_N^{} \bigr) 
      \Bigr] \, \gamma_{8}^{}   \;,  
\vspace{1ex} \\   \displaystyle   
\beta^{\prime(\rm S)}_{\Lambda p}  \;=&   
-\ratio{1}{\sqrt{6}}\, {\cal C}^2\, h_C^{}   
\;+\;  
\ratio{1}{\sqrt{6}}\, {\cal C}^2\, \Bigl[ 
-(m_\Lambda^{}-m_N^{}) + 3 \bigl( m_{\Sigma^*}^{}-m_N^{} \bigr) 
\Bigr] \, \gamma_{8}^{}   \;,    
\vspace{1ex} \\   \displaystyle   
\beta^{\prime(\rm S)}_{\Xi^-\Lambda}  \;=&   
\ratio{1}{\sqrt{6}}\, {\cal C}^2\, h_C^{}   
\;+\;  
\ratio{1}{\sqrt{6}}\, {\cal C}^2\, \Bigl[   
\ratio{4}{3} \bigl( m_\Xi^{}-m_\Lambda^{} \bigr)   
- 3 \bigl( m_{\Sigma^*}^{}-m_\Lambda^{} \bigr)   
- \ratio{16}{3} \bigl( m_{\Xi^*}^{}-m_\Lambda^{} \bigr)   
\Bigr] \, \gamma_{8}^{}   \;,    
\end{array}
\end{eqnarray*}
\begin{eqnarray*}         
\beta^{\prime(\rm P)}_{\Sigma^+ n}    &\!\!=&\!\!   \displaystyle     
{ -\ratio{1}{9} D\,{\cal C}^2 h_C^{} 
 + \Bigl( -\ratio{55}{162} {\cal H}+\ratio{8}{27} D-\ratio{4}{9} F \Bigr)  
  \, {\cal C}^2\, \bigl( h_D^{}-h_F^{} \bigr)  
 \over  m_\Sigma^{}-m_N^{} }  
\nonumber \\  && 
+\;   
{ \ratio{1}{3} D\,{\cal C}^2 h_C^{} 
 + \Bigl( \ratio{5}{162} {\cal H}-\ratio{4}{9} D-\ratio{4}{9} F \Bigr)  
  \, {\cal C}^2\, \bigl( h_D^{}+3 h_F^{} \bigr)  
 \over  m_\Lambda^{}-m_N^{} }   
\;+\;   
\Bigl( -\ratio{230}{243} {\cal H}   
      + \ratio{304}{81} D + \ratio{16}{9} F \Bigr)   
{\cal C}^2\, \gamma_{8}^{}   \;,   
\\ \nonumber \\   
\beta^{\prime(\rm P)}_{\Sigma^- n}    &\!\!=&\!\!   \displaystyle     
{ -\ratio{1}{9} F\,{\cal C}^2 h_C^{} 
 + \Bigl( \ratio{25}{54} {\cal H}-\ratio{26}{27} D+\ratio{2}{9} F \Bigr)  
  \, {\cal C}^2\, \bigl( h_D^{}-h_F^{} \bigr)  
 \over  m_\Sigma^{}-m_N^{} }  
\nonumber \\  && 
+\;   
{ \ratio{1}{3} D\,{\cal C}^2 h_C^{} 
 + \Bigl( \ratio{5}{162} {\cal H}-\ratio{4}{9} D-\ratio{4}{9} F \Bigr)  
  \, {\cal C}^2\, \bigl( h_D^{}+3 h_F^{} \bigr)  
 \over  m_\Lambda^{}-m_N^{} }   
\;+\;  
\Bigl( \ratio{170}{243} {\cal H}  
      + \ratio{82}{81} D + \ratio{22}{27} F \Bigr)   
{\cal C}^2\, \gamma_{8}^{}   \;,   
\\ \nonumber \\   
\beta^{\prime(\rm P)}_{\Lambda p}     &\!\!=&\!\!    \displaystyle     
{ \ratio{2}{9} D\, {\cal C}^2 h_C^{} 
 + \Bigl( -\ratio{5}{27} {\cal H}+\ratio{8}{3} D+\ratio{8}{3} F \Bigr) \, 
  {\cal C}^2\, \bigl( h_D^{}-h_F^{} \bigr)  
 \over  \sqrt{6}\; \bigl( m_\Sigma^{}-m_N^{} \bigr) }  
\nonumber \\  && 
-\;   
{ (D+F)\, {\cal C}^2 h_C^{} 
 + \Bigl( \ratio{10}{81} {\cal H}-\ratio{2}{3} D-\ratio{2}{9} F \Bigr) \, 
  {\cal C}^2\, \bigl( h_D^{}+3 h_F^{} \bigr)  
 \over  \sqrt{6}\; \bigr( m_\Lambda^{}-m_N^{} \bigr) }   
\;+\;  
\ratio{1}{\sqrt{6}} \Bigl( 
-\ratio{20}{27} {\cal H} + \ratio{106}{27} D + 6 F 
\Bigr) {\cal C}^2\, \gamma_{8}^{}   \;,      
\\ \nonumber \\   
\beta^{\prime(\rm P)}_{\Xi^-\Lambda}  &\!\!=&\!\!   \displaystyle     
{ \ratio{22}{9} D\, {\cal C}^2 h_C^{} 
 + \Bigl( \ratio{5}{27} {\cal H}-\ratio{8}{3} D-\ratio{8}{3} F \Bigr) \, 
  {\cal C}^2\, \bigl( h_D^{}+h_F^{} \bigr)  
 \over  \sqrt{6}\; \bigl( m_\Xi^{}-m_\Sigma^{} \bigr) }  
\nonumber \\  && 
-\;   
{ (D-F)\, {\cal C}^2 h_C^{} 
 + \Bigl( \ratio{20}{81} {\cal H}+\ratio{14}{27} D+2 F \Bigr) \, 
  {\cal C}^2\, \bigl( h_D^{}-3 h_F^{} \bigr)  
 \over  \sqrt{6}\; \bigl( m_\Xi^{}-m_\Lambda^{} \bigr) } 
\;-\;  
\ratio{1}{\sqrt{6}} \Bigl( 
\ratio{80}{81} {\cal H} + \ratio{58}{27} D + \ratio{58}{9} F 
\Bigr) {\cal C}^2\, \gamma_{8}^{}   \;.   
\end{eqnarray*}
%


The contributions from the wave-function renormalization of 
the pion and the octet baryons and from the renormalization of 
the pion-decay constant are collected into        
\begin{eqnarray*}         
\bar{\lambda}_{B_{}^{}B_{}'\pi}^{}  \;=\;  
\ratio{1}{2} \!\left( \bar{\lambda}_{B_{}^{}}^{} 
                   + \bar{\lambda}_{B_{}'}^{} 
                   + \lambda_\pi^{} \right) - \lambda_f^{} \;,       
\end{eqnarray*}    
where  $\,\bar{\lambda}_{B}^{}=\lambda_{B}^{}+\lambda_{B}',\,$  
$\lambda_\pi^{}$  and  $\lambda_f^{}$  are defined by   
\begin{eqnarray*}        
\Biggl( Z_{B}^{}, Z_{\pi}^{}, {f_{\!\pi}^{}\over f} \Biggr)  \;=\;  
1 + 
\Bigl( \bar{\lambda}_{B}^{},\lambda_{\pi}^{},\lambda_f^{} \Bigr) \,  
{m_K^2\over 16\pi^2 f_{\!P}^2}\, \ln{m_K^2\over\mu^2}   \;,  
\end{eqnarray*}    
with   
\begin{eqnarray*}         
\begin{array}{rlrl}   \displaystyle  
\lambda_{N}^{}  &\!\!=\;   
\ratio{17}{6} D^2 - 5 D F + \ratio{15}{2} F^2   \;, & \hspace{3em}   
\lambda_{N}'    &\!\!=\;   \ratio{1}{2} \,{\cal C}^2   \;, 
\vspace{1ex} \\   \displaystyle  
\lambda_{\Lambda}^{}  &\!\!=\;  \ratio{7}{3} D^2 + 9 F^2 
\;, & \hspace{3em} 
\lambda_{\Lambda}'    &\!\!=\;  {\cal C}^2   \;,  
\vspace{1ex} \\   \displaystyle  
\lambda_{\Sigma}^{}   &\!\!=\;  \ratio{13}{3} D^2 + 3 F^2 
\;, & \hspace{3em}   
\lambda_{\Sigma}'     &\!\!=\;  \ratio{7}{3} \,{\cal C}^2   \;,  
\vspace{1ex} \\   \displaystyle  
\lambda_{\Xi}^{}  &\!\!=\;   
\ratio{17}{6} D^2 + 5 D F + \ratio{15}{2} F^2   \;, \hspace{3em} &  
\lambda_{\Xi}'    &\!\!=\;  \ratio{13}{6} \,{\cal C}^2   \;,  
\vspace{1ex} \\   \displaystyle  
& \lambda_{\pi}^{}  \;=\;  -\ratio{1}{3}   \;, 
& \lambda_f^{}  \;=\;  -\ratio{1}{2}   \;.  & 
\end{array}   
\end{eqnarray*}    

One-loop corrections to the propagator that appears in tree-level 
pole diagrams in the P-waves yield the term 
$\tilde{\alpha}^{\rm(P)}_{B_{}^{}B_{}'}$.  
Its expression is equal to that of  $\alpha^{\rm(P)}_{B_{}^{}B_{}'}$   
with the exception that each factor  $1/(m_X^{}-m_Y)$  in   
$\alpha^{\rm(P)}_{B_{}^{}B_{}'}$  is replaced  with  
$\mu_{XY}^{}/(m_X^{}-m_Y^{})^2$,  where 
\begin{eqnarray*}   \label{muxy}
\mu_{XY}^{}  &\!\!=&\!\!      
- \Bigl( \bar{\beta}_{X}^{}-\bar{\beta}_{Y}^{} \Bigr)   
{m_K^3\over 16\pi f_{\!P}^2}   
+ \Bigl[ \Bigl( \bar{\gamma}_{X}^{} - \bar{\gamma}_{Y}^{} 
               - \bar{\lambda}_{X}^{}\alpha_{X}^{} 
               + \bar{\lambda}_{Y}^{}\alpha_{Y}^{} \Bigr) m_s^{}
        + \Bigl( \lambda_{X}' - \lambda_{Y}' \Bigr) \Delta m \Bigr]  
 {m_K^2\over 16\pi^2 f_{\!P}^2}\, \ln{m_K^2\over\mu^2}   \;, 
\label{mkcubed} 
\end{eqnarray*}      
with  
\begin{eqnarray*}         
\alpha_{N}^{}  =  -2 \left( b_D^{}-b_F^{} \right) - 2\sigma   
\;, \hspace{1em}   
\alpha_{\Lambda}^{}  =  -\ratio{8}{3} b_D^{} - 2\sigma   
\;, \hspace{1em}   
\alpha_{\Sigma}^{}  =  -2\sigma   
\;, \hspace{1em} 
\alpha_{\Xi}^{}  =  -2 \left( b_D^{}+b_F^{} \right) - 2\sigma   \;,
\end{eqnarray*}   
\begin{eqnarray*}         
\begin{array}{rl}   \displaystyle  
\bar{\beta}_{N}^{}  \;=&   
\ratio{5}{3} D^2 - 2 D F + 3 F^2   
+ \ratio{4}{9\sqrt{3}} \!\left( D^2 - 6 D F + 9 F^2 \right)   
\;+\;   
\ratio{1}{3} \,{\cal C}^2   \;,    
\vspace{1ex} \\   \displaystyle  
\bar{\beta}_{\Lambda}^{}  \;=&   
\ratio{2}{3} D^2 + 6 F^2 + \ratio{16}{9\sqrt{3}} D^2   
\;+\;   
\ratio{2}{3} \,{\cal C}^2   \;,   
\vspace{1ex} \\   \displaystyle   
\bar{\beta}_{\Sigma}^{}  \;=&   
2 D^2 + 2 F^2 + \ratio{16}{9\sqrt{3}} D^2 
\;+\;
\left( \ratio{10}{9} + \ratio{8}{9\sqrt{3}} \right) {\cal C}^2   \;, 
\vspace{1ex} \\   \displaystyle  
\bar{\beta}_{\Xi}^{}  \;=&   
\ratio{5}{3} D^2 + 2 D F + 3 F^2 
+ \ratio{4}{9\sqrt{3}} \!\left( D^2 + 6 D F + 9 F^2 \right) 
\;+\;  
\left( 1 + \ratio{8}{9\sqrt{3}} \right) {\cal C}^2   \;,     
\end{array}   
\end{eqnarray*}   
\begin{eqnarray*}   \label{endmassr}   
\begin{array}{rl}   \displaystyle  
\bar{\gamma}_{N}^{}  \;=&   
\ratio{43}{9}\, b_D^{} - \ratio{25}{9}\, b_F^{} 
- b_D^{} \left( \ratio{4}{3} D^2 + 12 F^2 \right)  
+ b_F^{} \left( \ratio{2}{3} D^2 - 4 D F + 6 F^2 \right)  
+ \ratio{52}{9}\, \sigma - 2 \sigma \lambda_N^{}     
\vspace{1ex} \nonumber \\  & 
+\;   
\ratio{1}{3}\, c \,{\cal C}^2 - 2 \tilde{\sigma} \lambda_N'   \;,   
\vspace{2ex} \\   \displaystyle   
\bar{\gamma}_{\Lambda}^{}  \;=& 
\ratio{154}{27}\, b_D^{} 
- b_D^{} \left( \ratio{50}{9} D^2 + 18 F^2 \right)  
+ b_F^{} \left( 12 D F \right)  
+ \ratio{52}{9}\, \sigma - 2 \sigma \lambda_\Lambda^{}     
\;+\; 
\ratio{4}{3}\, c \,{\cal C}^2 - 2 \tilde{\sigma} \lambda_\Lambda'   \;,   
\vspace{2ex} \\   \displaystyle   
\bar{\gamma}_{\Sigma}^{}  \;=&   
2 b_D^{} 
- b_D^{} \left( 6 D^2 + 6 F^2 \right)  
- b_F^{} \left( 12 D F \right)  
+ \ratio{52}{9}\, \sigma - 2 \sigma \lambda_\Sigma^{}   
\;+\;    
\ratio{8}{9}\, c \,{\cal C}^2 - 2 \tilde{\sigma} \lambda_\Sigma'   \;,  
\vspace{2ex} \\   \displaystyle   
\bar{\gamma}_{\Xi}^{}  \;=&   
\ratio{43}{9}\, b_D^{} + \ratio{25}{9}\, b_F^{} 
- b_D^{} \left( \ratio{4}{3} D^2 + 12 F^2 \right)  
- b_F^{} \left( \ratio{2}{3} D^2 + 4 D F + 6 F^2 \right)  
+ \ratio{52}{9}\, \sigma - 2 \sigma \lambda_\Xi^{}   
\vspace{1ex} \nonumber \\  & 
+\;   
\ratio{29}{9}\, c \,{\cal C}^2 - 2 \tilde{\sigma} \lambda_\Xi'   \;.     
\end{array}
\end{eqnarray*}

\end{document}